\input epsf
\documentstyle[prd,aps]{revtex}
\def\ut#1{\rlap{\lower1ex\hbox{$\sim$}}#1{}}
\def\uti#1{\rlap{\lower1ex\hbox{$\scriptstyle \sim$}}#1{}}
\begin{document}

\title{The space of states of quantum gravity in terms of loops \\and extended 
loops: some remarks}

\author{Cayetano Di Bartolo$^1$, Rodolfo Gambini$^2$, Jorge Griego$^2$
and Jorge Pullin$^3$}
\address{1. Departamento de F\'{\i}sica, Universidad Sim\'on Bol\'{\i}var,
Caracas, Venezuela\\
2. Instituto de F\'{\i}sica, Facultad de Ingenier\'{\i}a, J. Herrera y Reissig
565, Montevideo, Uruguay\\
3. Center for Gravitational Physics and Geometry, Department of Physics, \\
104 Davey Lab, The Pennsylvania State
University, University Park, PA 16802}

\date{\today}
\maketitle

\begin{abstract}

This article reviews the
status of several solutions to all the constraints of quantum gravity
that have been proposed in terms of loops and extended loops. We
discuss pitfalls of several of the results and in particular discuss
the issues of covariance and regularization of the constraints in
terms of extended loops. We also propose a formalism for ``thickened
out loops'' which does not face the covariance problems of extended
loops and may allow to regularize expressions in a consistent manner.
\end{abstract}

\vspace{-9.5cm} 
\begin{flushright}
\baselineskip=15pt
CGPG-95/3-5  \\
gr-qc/9503059\\
\end{flushright}
\vspace{9cm}

\section{Introduction}

The intention of this paper is to review the state of the art in
February 1995 of what is perceived as one of the main achievements of
the new variable canonical quantization program: the possibility to
find in a generic case solutions to the Wheeler-DeWitt equation of
quantum gravity. This has been one of the main motivations for the
study of the new variables and led to many of its most striking
results: the introduction of holonomies  \cite{JaSm},
the loop representation and its connections with knot theory
\cite{RoSm88}, the connection with Chern-Simons theory and the Jones
polynomial \cite{BrGaPu93} and to the formulation of the extended
loop representation \cite{DiGaGr}.  Due to the dynamic character of
research in this field many results have superseded or changed the
perspective on previous ones and a review of the different problems is
in order. We will try to structure this article chronologically and
discuss many new aspects of old results that have been discovered with
subsequent developments.

For reasons of space, this article is not self-contained. For the sake
of brevity we will try to concentrate on outstanding issues, and
referring the reader to appropriate references for details.  The
structure of this article is as follows: in section 2 we will discuss
the Hamiltonian and the formulation of the theory in terms of
loops. We will then discuss solutions in terms of loops
based on their intersection structure and
solutions based on the Jones polynomial and issues of their
regularization. In section 4 we will briefly introduce the extended
representation as a solution to the problem of regularizing
wavefunctions and study the issue of the regularization and
renormalization of the constraints. In section 5 we discuss the 
existence of regularized solutions and the need for counterterms in the 
renormalized constraints. In section 6 we will discuss the
problem of the gauge covariance of the extended representation and
possible solutions based on thickened out loops.

\section{The loop representation}

Whenever one has a theory given in terms of a Lie-algebra valued 
connection $A_a^i$ on a three dimensional manifold, one can introduce 
a loop representation for it. One way of doing this is via the loop 
transform,

\begin{equation}
\Psi(\gamma) = \int d A W_\gamma[A] \Psi[A]
\label{transform}
\end{equation}
where $W_\gamma[A]$ (``Wilson loop'') is the trace of the holonomy of
the connection $A$ along the loop $\gamma$. The transform is a
functional integral and we refer to the article by Ashtekar et al
\cite{As95} in this volume for progress concerning its rigorous
definition. Here we will use it in a heuristic way.

It will be convenient to view wavefunctions in the loop
representation $\Psi(\gamma)$ as functions on the group of loops
base-pointed at a point $o$, ${\cal L}_o$ \cite{GaTr81}. Because of
the definition (\ref{transform}) these wavefunctions satisfy a
series of identities. To begin with they are base-point-independent
because the trace of a holonomy is. As a consequence of the properties
of products of traces of matrices in a Lie-group one has the
Mandelstam identities, which for the case of interest ($SU(2)$) are
given by, 
\begin{eqnarray} \Psi(\gamma_1 \circ \gamma_2) &=&
\Psi(\gamma_2 \circ \gamma_1) \\ \Psi(\gamma) &=&\Psi(\gamma^{-1})\\
\Psi(\gamma_1\circ\gamma_2\circ\gamma_3)
+\Psi(\gamma_1\circ\gamma_2\circ\gamma_3^{-1})
&=&\Psi(\gamma_2\circ\gamma_1\circ\gamma_3) +
\Psi(\gamma_2\circ\gamma_1\circ\gamma_3^{-1}) \label{mandelfund}\end{eqnarray}

Some comments are in order. Sometimes loop representations have been
constructed based on multi-loops. In that case one considers in the
transform a product of Wilson loops, one for each loop forming the
multi-loop.  If one proceeds that way, there are also Mandelstam
identities relating wavefunctions of different multi-loops. It turns
out that for the $SU(2)$ case these identities allow to express all
multi-loop wavefunctions in terms of functions of single loops. We will
assume such reduction has been performed and work with a single loop
from now on. Notice that even with a single loop one has identities,
the ones we listed above. By suitable combinations of these one can
get an infinite number of identities that must be satisfied by the
wavefunctions. A procedure to characterize a set of independent loops
under these identities is described in the article by Rovelli and
Smolin in this volume \cite{RoSm95}. There are other properties
satisfied by traces of holonomies, for instance inequalities
\cite{GaTr86,Lo93}, that reflect properties of the gauge group. It is not
clear at present what are the consequences on the space of wavefunctions
of there properties, so we will ignore them here. These and other
problems are intimately related to the precise definition of a loop
transform, as can be seen in particular examples \cite{Ma93} where the
transform can be explicitly computed and in current efforts to define
rigorously the transform. 

Viewing the wavefunctions as functions of the group of loops has the
advantage of immediately incorporating another symmetry of the
wavefunctions, the invariance under the addition of retraced paths of
vanishing area (``trees''), $\Psi(\alpha) = \Psi(\alpha\circ \eta_x
\circ \eta^{-1}_x)$ where $\eta_x$ is an open path going from some
point on the loop to the point $x$. This invariance, coupled with the 
Mandelstam identities, constrains the kind of functions and operators one
can consider in the loop representation. Any wavefunction not complying
with these identities is unacceptable, no matter what its behavior may
be with respect to the operators of the theory. 

An example of such a function, that had historical relevance, is the 
``characteristic'' function of smooth loops, defined by,
\begin{equation}
\Psi(\gamma)=\left\{\begin{array}{cc}1 & \mbox{if $\gamma$  is
smooth and non-intersecting}\\0& \mbox{otherwise}\end{array}\right.
\end{equation}

This function is apparently annihilated by the Hamiltonian constraint,
as we will see shortly. However, such a function is not acceptable as
a wavefunction of the gravitational field because it is not compatible
with the Mandelstam identities. To see this, consider a smooth loop as
a composition of three loops, as shown in figure 1,
$\gamma=\gamma_1\circ\gamma_2\circ \gamma_3$ and apply the Mandelstam
identity (\ref{mandelfund}). It will give a combination of loops with
intersections. One quickly runs into an inconsistency since
$\psi(\gamma)$ is one since $\gamma$ is non-intersecting and has to be
equal to something vanishing \cite{Asrecent}. It may be possible,
through the introduction of a non-trivial inner product to construct
states compatible with the Mandelstam identity and related to
non-intersecting loops, see \cite{GaPubook,Smgift}. If such states are
to be annihilated by the Hamiltonian constraint, it should be
a self-adjoint operator with respect to the inner product introduced.

\begin{figure}
\hspace{2cm}\epsfxsize=200pt \epsfbox{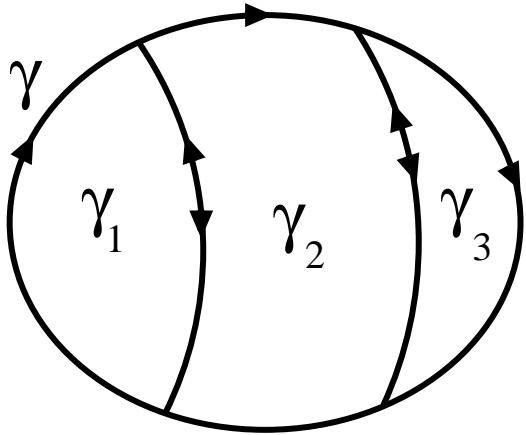}
\caption{The loop used in the Mandelstam identity that is 
not satisfied by the naive states}
\label{naivemal}
\end{figure}

Using the formal transform (\ref{transform}) one can obtain an expression in 
the loop representation for the constraints of general relativity. The 
expressions one gets are the following,

\begin{eqnarray}
{\cal C}(\vec{N}) \Psi(\gamma)&=& 
\int d^3x N^a(x) \oint_\gamma dy^b \delta(x-y) 
\Delta_{ab}(\gamma_o^y) \Psi(\gamma)\\
{\cal H}(\ut{M})(x) \Psi(\gamma)&=& 
\int d^3x \ut{M} \oint_\gamma dy^a 
\oint_\gamma dz^b f_\epsilon(x-z) \delta(z-y) \Delta_{ab}(\gamma_o^z) 
\Psi(\gamma_y^z\circ \gamma_{yo}^z)
\end{eqnarray}

These expressions have been analyzed in detail in the past 
\cite{Ga91,GaPubaez,GaPubook} so we will not discuss them here. 
It has also been
shown \cite{BrPu93} that they are equivalent to the original ones 
introduced by Rovelli and Smolin \cite{RoSm90}. We would like to make a few
comments about them.

The first remark is that if the loop is smooth at the point in which
the Hamiltonian acts, then its action is zero (we are ignoring issues
related to the ``acceleration terms'', which are irrelevant for
diffeomorphism invariant states see \cite{GaPubook,Bo95}). That is the
reason why the characteristic states introduced above were
automatically annihilated.

  The expressions we give for the constraints are written in terms of
loop derivatives. The loop derivative is an operator that acts on the
space of functions of base-pointed loops. Its action per se is
therefore not well defined on $\Psi(\gamma)$'s, which are not
base-pointed. It is the fact that the loop derivative appears under a
loop integral that makes the expression for the constraints well
defined. The expressions are computed fixing a fiducial base-point in
the loop and the result is independent of this choice. The first
constraint has been known for some time to be the generator of
infinitesimal deformations in loop space. Solutions to the first
constraint are functions of loops invariant under deformations of the
loops, that is, they are knot invariants. At this point another puzzle
appears. Evidently the definition of the above constraints requires
that the wavefunctions considered be loop differentiable. That means
that they should change smoothly under the addition of an
infinitesimal loop. An example of such functions would be holonomies
of smooth connections. Knot invariants, however, are not in general
loop differentiable. A knot invariant does not behave smoothly under
deformations of the loops, its value jumps discontinuously if the
deformation changes the topology of the loop. The loop
derivative can do this and therefore knot invariants are in general
not loop differentiable.  The expressions of the constraints we
introduced above will therefore require further elaboration if we want
them to act on knot invariants.

The intuitive picture that therefore arises is that the diffeomorphism
constraint requires us to work with wavefunctions that viewed as
functions on loop space behave as ``step functions''; they change
values abruptly when one goes from one knot class to another. The only
hope to define the action of differential operators on them in any
consistent fashion is through a regularization. Notice that this is a
regularization that appears {\em in addition} to the usual need
to regularize the quantum operators in a field theory. It is a
regularization at the level of wavefunctions.

Let us discuss these issues with some concrete examples. There exist in
the literature several knot invariants for which explicit analytic
expression in terms of loops can be given. These are the only
invariants amenable to a concrete calculation for the action of the
differential operators on them. Unfortunately none of these invariants
is well defined for arbitrary loops (especially if one includes 
intersections). Some of these invariants are in fact not even well
defined for ordinary smooth loops. An example of this is provided by
the Gauss linking number of a loop with itself (the self-linking
number). An analytic expression for the self-linking number is,

\begin{equation}
L(\gamma) = {1\over 4\pi} \oint_\gamma dy^a \oint_\gamma dz^b
\epsilon_{abc}{(y-z)^c\over
|y-z|^3} \label{gauss}
\end{equation}

in extended loop notation it would be written as,

\begin{equation}
L(\gamma) =   -2 g_{ax\,by} X^{ax\,by}(\gamma)
\end{equation}
where $g_{ay\,bz} =-1/(4\pi) \epsilon_{abc} {(y-z)^c\over |y-z|^3} $ is the
Green function (propagator) of a Chern-Simons theory and $X^{ax\,by}$
is the second order multi-tangent associated with a loop $\gamma$
\cite{DiGaGr}. It is easy to see that as a function of a loop 
it satisfies the Mandelstam identities and therefore it is a 
candidate wavefunction for quantum gravity.

The expression (\ref{gauss}) would be a genuine link invariant if the
first and second integrals were evaluated on different loops. It would
just count the linking number of both loops. In particular all
information about the fiducial metric introduced to compute $|x-y|$
would be erased. When one evaluates the linking number of a loop with
itself, the expression has a problem when $y=z$. In fact one can see
\cite{Ca} that the result of the integral is finite, but is dependent
on the particular choice of fiducial metric performed to compute the
Chern-Simons propagator. This is an example of what we meant by some
knot invariants not being well defined even for smooth loops. If the
loop has an intersection there is an added difficulty, since the
resulting expression is not even finite.

A proposed solution to this problem is to consider the invariant
defined by equation (\ref{gauss}) as an invariant of a {\em framed}
loop. A framing is a procedure that given a loop assigns another loop
infinitesimally separated from the original one. Another way of
putting this is that it associates a preferred ``ribbon'' or a normal
vector to a loop. The self-linking number for the framed loop is
simply defined as the linking number of the two loops in the
ribbon. For the self-linking number such a procedure also removes
singularities in the case of intersections, since the two integrals
are simply evaluated on different loops and no singularities
appear. This situation is generic. Even for invariants that do not
require a framing when evaluated on smooth loops (as is the case of
the second coefficient of the Conway polynomial defined in
\cite{GuMaMi}) one requires some procedure to remove singularities
generated by intersections. We call all these procedures
``framings''. (More discussion on framings can be found in references
\cite{Wi89,GuMaMi}).

How would one proceed to define the action of the Hamiltonian
constraint introduced above on invariants as the ones we have been
discussing? Two possibilities immediately come to mind, associated
with taking in different orders the two limits involved in defining
the framing and the loop derivative: 1) To act with
the loop derivative on the invariant evaluated for a framed loop,
for a finite separation of the framing; 2) To act with the loop
derivative on a framed loop in such a way that the action of the derivative
could change the relative knotting of the two loops in the framing.

In the first case the action of the loop derivative is trivial. It
acts on each loop individually and does not modify their relative
linking. If one were to pursue this seriously any invariant would be
annihilated by all the constraints. This does not seem a sensible
choice, since there is no reason why any knot invariant should be
annihilated by the Hamiltonian constraint. The other possibility, in
which the loop derivative can change the relative topology of both
loops in the framing has the disadvantage of being ill defined. The
problem is that one could imagine situations in which the result of
the action of the derivative would be dependent on the particular
details of the loop appended in order to compute the loop
derivative. It is convenient to recall that in the definition of the
loop derivative, the shape of the loop did not matter as long as it
spanned a certain area. This is not the case for the action of the
Hamiltonian we obtain.

Is this just a problem of these particular wavefunctions or is it
generic? What happens if one considers a function of loops that is
well defined without introduction of a framing? As we mentioned above,
the loop derivative is simply not defined. Is this therefore a problem
of the particular way of writing the constraint in terms of the loop
derivatives? Could one rearrange limits in such a way that the
Hamiltonian has an action that is well defined on knots? A recent
paper \cite{RoSm94} has explored these and other issues. The
conclusion of that particular analysis is that one can (ignoring
certain subtleties \cite{Bo95}) find an operator that reduces in the
connection representation to the Hamiltonian constraint and that has
an action on knots. The trouble is that the action inherits one of the
difficulties mentioned above. It corresponds to the addition of a
small loop and the result therefore depends on how one adds that
loop. One started with an operator that did not know about this small
loop and ends up with a result that is dependent on it. It may be the
case that this is inevitable and is just a reflection of the
infinitely many choices one has when defining a regularization of an
operator in a quantum field theory. At the moment the status of this problem
is unclear.

Here we will explore yet another option. We will consider knot
invariants given by expressions similar to the self-linking number we
considered above. We will study the action of the Hamiltonian defined
in the following way. Given an invariant generically written as ``$g
\cdot X(\gamma)$'' where $g$ is a set of propagators and $X(\gamma)$ 
is a multi-tangent of arbitrary order, we will act with the Hamiltonian
on $X(\gamma)$ and compute the result. Notice that the loop derivative
of the $X(\gamma)$ is well defined, so this seems to bypass the
problem of differentiating a knot invariant. This is not the case
since the contraction with the $g$'s and the resulting action of the
Hamiltonian are singular. The calculations we will perform are
therefore purely formal.  As a result we will find that several quite
nontrivial knot invariants are annihilated formally by the Hamiltonian
constraint. The extended loop representation which we discuss in the
next section will provide an arena in which completely equivalent
results appear in a fully regularized context.

There is a result in the connection representation that is the key to
all the solutions we will discuss. As was first observed by Kodama
\cite{Ko}, if one considers a wavefunction given by the exponential of
the Chern-Simons form of the Ashtekar connection, 

\begin{equation}
\Psi_{\Lambda}[A] = {\rm exp}(-{\textstyle{6 \over \Lambda}}
S_{CS})\equiv
{\rm exp} (-{\textstyle{6 \over \Lambda}}
\int \tilde{\eta}^{abc} Tr[A_a
\partial_b A_c +{\textstyle 2 \over 3} A_a A_b A_c])
\end{equation}
it is a solution of all the constraints of quantum gravity with a
cosmological constant $\Lambda$ (there is a factor ordering issue with
respect to the diffeomorphism constraint that was solved in
\cite{BrGaPu93}).  If one considers the transform of this state, one
obtains an integral

\begin{equation}
\Psi[\gamma] = \int d A\quad W_\gamma[A]
e^{-{\textstyle{6 \over \Lambda}} S_{CS}},
\label{lpcs}
\end{equation}
that formally looks the same as the expectation value of a Wilson loop
in a Chern-Simons theory. This expression has been studied by many
authors \cite{Wi89,Sm88,GuMaMi} and the result is that it is a knot
invariant that is known as the Kauffman bracket knot polynomial. Since
the integral (\ref{lpcs}) can be computed using perturbation theory
(since Chern-Simons theory is renormalizable) one actually can obtain
concrete expressions for the different coefficients of the Kauffman
bracket. 

If one believes these results one is forced therefore to conclude that
the Kauffman bracket should be annihilated by the Hamiltonian
constraint of quantum gravity with cosmological constant in the loop
representation. This is a calculation that can be checked, at least at
the formal level discussed above. The main result is that it is in
fact a solution and as a surprising consequence of the calculation one
finds that certain portion of the coefficients of the Kauffman
bracket, which coincides precisely with a coefficient of the Jones
polynomial, are annihilated by Hamiltonian constraint with
$\Lambda=0$. The Kauffman bracket is known to be related to the Jones
polynomial (in the arbitrary variable $q$) through the expression,

\begin{equation}
{\rm Kauffman}(\gamma)_q = q^{{3 \over 4} L(\gamma)} 
{\rm Jones}(\gamma)_q,
\end{equation}
which highlights the fact that the framing dependence of the Kauffman
bracket is concentrated in the prefactor involving the self linking
number, since the Jones polynomial is framing independent.

Unfortunately, an initial conjecture that suggested that all the
coefficients of the Jones polynomial were solutions turned out not to
be true. Detailed calculations show that the third coefficient of the
Jones polynomial is not annihilated by the vacuum Hamiltonian
constraint. This is in spite of the fact that the Kauffman bracket
{\em is} annihilated up to that order.  In pursuing this analysis it
was also noted \cite{GaPubaez} that at the root of the result is the
fact that the prefactor is formally a solution of the Hamiltonian
constraint with cosmological constant. It can be checked that all
these states satisfy the Mandelstam identities and solve the
constraints without any assumption about the loops (they can have
an arbitrary number of intersections).

In spite of the fact of the nontrivial nature of all these results,
which arise due to elaborate cancelations of terms in the action of
the Hamiltonian, we remind the reader that the results are all formal
in the sense discussed above. This non-triviality of the results is the
main motivation to seek a framework that would allow to make sense of
them in a fully regularized context.

\section{The extended loop representation}

The extended loop representation is obtained by replacing the
multi-tangents (also called ``form factors'') that appear in the
definition of the holonomy by arbitrary multi-vector densities. What
is the point of doing this? Consider the expression we introduced in
the previous section for the self linking number. If instead of being
given by a double loop integral it had been given by a double spatial
integral along smooth vector densities $X^a(y)$,
\begin{equation}
L(X) ={1 \over 4 \pi} 
\int d^3y \int d^3z X^{a}(y) X^{b}(z) \epsilon_{abc} {(y-z)^c
\over |y-z|^3}
\end{equation}
the integral would have been well defined. The above integral yields
the usual expression if one replaces the smooth vector densities by
the distributional tangents to the loops $X^a(y)(\gamma) =
X^{ay}(\gamma) \equiv \oint_\gamma dz^a \delta(z-y)$. The problem we are
confronted with is analogous to the usual difficulty in electrostatics: if
one computes the energy of a set of distributional charges a
divergence appears if the charges touch. If the distributions are
smooth no problem arises in spite of the fact that the propagator is
distributional. 

The rationale for extended loops is therefore to replace the ill
definitions that arise in terms of loops due to their distributional
nature by replacing the information needed about the loops tangents by
smooth fields. Doing this in the non-Abelian case requires some
detailed work that we briefly discuss. More complete treatments can be
found in \cite{DiGaGr,DiBaGaGr}.

The starting point is the expression for the ``extended holonomy'', 
\begin{equation}
U({\bf X}) = {\bf X} \cdot {\bf A} \equiv \sum_{n=0}^\infty
X^{\uti{\mu}_n} A_{\uti{\mu}_n}
\end{equation}
where $X^{\uti{\mu}_n}$ is a shorthand for
$X^{a_1\,x_1,\ldots,a_n\,x_n}$ and $A_{\uti{\mu}_n}$ for
$A_{a_1}(x_1)\cdots A_{a_n}(x_n)$ and $X$ is a set of multivector
densities. If one would like to recover the usual holonomy of a loop
one needs to replace the $X$'s by appropriate multi-tangents to the
loop, details are given in \cite{DiGaGr}.

Obviously such an expression does not in general converge. It does not
seem reasonable to discuss convergence criteria for it, since in
general it is not even gauge covariant.  If one requires (formally)
that the extended holonomy have a gauge invariant trace one needs to
require certain conditions on the multi-tensors,
\begin{eqnarray}
&&\frac {\partial \phantom{iii}} {\partial x_i^{a_i}} \,
X^{a_1 x_1\ldots a_i x_i \,\ldots\,a_nx_n} =\label{dc}
\\
&&\hspace{1cm} \bigl( \,\delta(x_i-x_{i-1}) - \delta(x_i-x_{i+1}) \,\bigr)
 X^{ a_1 x_1\ldots a_{i-1} x_{i-1}\,a_{i+1} x_{i+1}\ldots a_n x_n}{}.\nonumber
\end{eqnarray}
which are usually referred to as ``the differential
constraint''. In this expression, points $x_0$ and $x_{n+1}$ are to be
understood as the base-point of the loop. The multi-tangents of a usual
loop satisfy this equation.

Now that we know how to obtain gauge invariants from extended loops,
can we answer in a meaningful way the question of the convergence of
the extended holonomy? The answer is again negative. Although the
trace of the whole series is gauge invariant, each individual term is
not. The usual way to determine if a series converges is by comparing
successive terms in the series. Such criteria therefore cannot be
translated into a gauge invariant statement. We will therefore simply
have to assume that the extended loops we are dealing with are those
for which the extended holonomy converges for the portion of the space
of connections modulo gauge transformations of physical interest.

Even assuming this, in which sense does the differential constraint ensure
that the trace of the extended holonomy is gauge invariant? If one
performs an infinitesimal gauge transformation of the extended
holonomy one gets an infinite series. The differential constraint
ensures that terms in that series cancel in pairs. Of course, this
only implies gauge invariance given certain properties of convergence
of the series. The precise statement is that the term \cite{Sch95},

\begin{equation}
\sum_{k=1}^N A_{\mu_1}\ldots A_{\mu_{k-1}} [A,\Lambda]_{\mu_k} A_{\mu_{k+1}}
\ldots A_{\mu_n}X^{\mu_1\,\ldots\,\mu_n}\label{conv}
\end{equation}
should go to zero as $N\rightarrow\infty$. $\Lambda$ is the parameter
of the gauge transformation and is therefore an arbitrary
function. Notice that the vanishing of (\ref{conv}) is not guaranteed
by the convergence of the holonomy alone. 

These problems of convergence were illustrated in a recent paper by 
by Schilling \cite{Sch95} where explicit examples of extended loops
that satisfied the differential constraint and that failed to give
extended holonomies with invariant trace were constructed. Does this
mean that the formalism is doomed? We believe the answer is no. 
We will show in section 5 that one
can actually find a large subset of extended loops for which the
convergence problems do not arise and for which the following
framework would be applicable.

Another comment about the differential constraint is that due to its
structure it mandates that the multi-tangents should be distributional
objects. Does this defeat the whole purpose of extended loops? After
all, we introduced them in order to smoothen out the distributional
behavior of ordinary loops. We will see that the following set of
ideas allows us to make use of these objects, in spite of the fact
that they are distributional.

The $X$'s that satisfy (\ref{dc}) form a group with composition law,
\begin{equation}
({\bf X}_1\times{\bf X}_1)^{\uti{\mu}_n} \equiv \sum_{k=1}^n 
X_1^{\mu_1,\ldots,\mu_k} X_2^{\mu_{k+1},\ldots,\mu_n}, 
\end{equation}
where $\mu_i$ stands for a paired vector index and spatial point
$a_i,x_i$. This group is called the extended group of loops. The
ordinary group of loops arises in the limit when one replaces $X$'s by
the multi-tangents to a loop. Associated with the extended group is an
infinite dimensional Lie algebra. Due to this one can write a generic
element of the extended group of loops as,
\begin{equation}
{\bf X} = \exp(\mbox{\Large$\sigma$\normalsize}\cdot {\bf Y}),
\end{equation}
where the exponential is determined by the product introduced above,
the {\mbox{\Large$\sigma$\normalsize}'s are a basis for the
algebra given explicitly in \cite{DiGaGr} and $Y$ is an {\em
arbitrary} multitensor. We omit all details. The point is that because
the $Y$'s are arbitrary, one in particular can choose them to be
smooth. As a consequence of this we see that all the distributional
structure of the $X$'s has to arise from the
{\mbox{\Large$\sigma$\normalsize}'s. Since these are known explicitly,
this structure is well under control. We will see that this explicit
knowledge of the singularity structure is a powerful aid in
regularizing the wavefunctions and the action of the constraints on
them. 

In fact if one considers the wavefunctions given by the knot
invariants explicitly introduced in the previous section from
Chern-Simons theory, the ones that looked like ``${\bf g}\cdot {\bf
X}$'', it can be explicitly checked \cite{DiBaGaGr} that the particular
form of the ${\bf g}'s$ that comes from Chern-Simons theory is such
that if one introduces the $X$'s stemming from smooth $Y$'s introduced
above, all the divergent structure is annihilated and one is left with
a well defined smooth expression purely given in terms of the $Y$'s.
This is due to the particular form and symmetries of the propagators that
appear for these wavefunctions contracted with the $X$'s. 
Therefore we see that the extended loops have delivered their promise:
due to their less singular nature than loops they manage to provide
expressions for the gauge invariants that are well defined.

In order to write expressions for the constraints in the extended
representation it is worthwhile to notice that due to the Mandelstam
identities wavefunctions only depend on the extended loops through the
combination,
\begin{equation}
R^{\mu_1 \ldots \mu_n} = \frac{1}{2} \left[  X^{\mu_1
\ldots  \mu_n}  + (-1)^n X^{\mu_n \ldots \mu_1} \right].
\end{equation}

Acting on functions of $R$ the constraints can be written as linear
functional operators (for details see \cite{DiBaGaGr}),

\begin{eqnarray}
{\cal C}(\vec{N}) \Psi({\bf R}) &=& \int d^3x N^a(x) ({\cal F}_{ab}(x) \times
{\bf R}^{(bx)})^{\uti{\mu}} {\delta \over \delta {\bf R}^{\uti{\mu}}}
\Psi({\bf R}) \\
{\cal H}(\ut{M}) &=& 2
\int d^3 x \ut{M}(x) ({\cal F}_{ab}(x)\times {\bf R}^{(ax,bx)})^{\uti{\mu}}
{\delta \over \delta {\bf R}^{\uti{\mu}}} \Psi({\bf R})
\end{eqnarray}
where ${\cal F}_{ab}^{\uti{\mu}}(x)$ is an element of the extended algebra such
that ${\cal F}_{ab}^{\uti{\mu}}(x) {\bf A}_{\uti{\mu}} = {\bf F}_{ab}$,
$(R^{(bx)})^{\uti{\mu}}\equiv R^{(bx\,\uti{\mu})_c}$, the notation $()_c$
meaning cyclic permutation of indices. The element 
${\bf R}^{(ax,bx)}$ is given by
\begin{equation}
[{\bf   R}^{(ax, \,bx)}]^{\uti{\rho}} =
R^{ (ax, \,bx) \uti{\rho}} \equiv
({\bf \delta}_{\uti{\nu}} \times
{\bf \delta}_{\uti{\mu}})^{\uti{\rho}}\,(-1)^{n({\uti{\mu}})}\,
R^{(ax\, \uti{\nu} \,bx\,\uti{\mu}^{-1})_c}.
\label{rsymmetric}
\end{equation}

It is remarkable that the operators have such a similar action, though
the notation helps to conceal important differences. They are basically
concentrated in the special elements with ``marked'' points $x$ that
accompany the ${\cal F}_{ab}$ in the expression of the constraints. The
element ${\bf R}^{(ax,bx)}$ is evidently divergent since it has a
spatial dependence repeated and we know through the differential
constraint that the object is distributional. That is the reason why the
Hamiltonian still requires a regularization in spite of being a linear
operator. A point-splitting regularization is simply achieved by
separating the two $x$'s in the $R^{(ax,bx)}$.

As we discussed above, the wavefunctions we introduced in the previous
section are simply generalized to the extended representation by
formally substituting the dependence on the multi-tangents to a loop by
the dependence on an arbitrary multitensor stemming from a smooth ${\bf
Y}$. Let us exhibit this explicitly for the second coefficient of the
Alexander-Conway polynomial. Its expression in terms of multi-tangents
fields is \cite{GuMaMi},

\begin{equation}
a_2[\gamma]  = h_{\mu_1 \mu_2 \mu_3}
X^{\mu_1 \mu_2 \mu_3}(\gamma) +
g_{\mu_1 \mu_3} g_{\mu_2 \mu_4} X^{\mu_1 \mu_2 \mu_3 \mu_4}(\gamma)
\end{equation}
where
\begin{equation}
h_{\mu_1 \mu_2 \mu_3} = \int d^3 z \epsilon^{b_1 b_2 b_3} \;
g_{\mu_1\,b_1 z} \, g_{\mu_2\,b_2 z}\, g_{\mu_3\,b_3 z}
\end{equation}
with 
\begin{equation}
\epsilon^{c_1 z_1\,c_2 z_2\,c_3 z_3} = \epsilon^{c_1 c_2 c_3}
\int d^3 t \; \delta(z_1 -t)\,\delta(z_2 -t)\,\delta(z_3 -t)
\end{equation}
The generalization of this knot invariant to extended loops is
straightforward
\begin{equation}
a_2[\gamma] = a_2[{\bf X}(\gamma)]\rightarrow a_2({\bf X}) = a_2({\bf R}).
\end{equation}

We now analyze the application of the Hamiltonian constraint on this
state in the extended representation. With the expression introduced
above we have,
\begin{eqnarray}
&& \hspace{-1cm} {\cal  H}(x) \,a_2 ({\bf R} )  =
2\,h_{\mu_1\mu_2\mu_3}
\left[ {\cal F}_{ab}{^{\mu_1}} (x) \, R^{(ax,\,
bx) \mu_2 \mu_3}
+ {\cal F}_{ab}{^{\mu_1 \mu_2}} (x) \, R^{(ax,\,
bx) \mu_3 } \right]  \nonumber \\
&&
+ 2\,  g_{\mu_1\mu_3}g_{\mu_2\mu_4}
\left[ {\cal F}_{ab}{^{\mu_1}} (x) \, R^{(ax,\,
bx) \mu_2 \mu_3 \mu_4 }
+ {\cal F}_{ab}{^{\mu_1 \mu_2}} (x) \, R^{(ax,\,
bx) \mu_3 \mu_4 } \right]
\end{eqnarray}

We can compute the action of ${\cal F}_{ab}$ on  the  propagators $g$
by integration by parts and the use of their explicit expressions. The
results are,
\begin{eqnarray}
\lefteqn{ \hspace{-0.5cm} 
 {\cal F}_{ab}{^{\mu_1 }}(x) \, g_{\mu_1 \mu_3} = -
\epsilon_{ab a_{3}} \delta(x-x_3)
- \partial{_{a_{3}}} g_{ax \, bx_{3}}} \label{R1sobreg} \\
\lefteqn{ \hspace{-0.5cm} 
{\cal F}_{ab}{^{\mu_1 \mu_2}}(x) \, g_{ \mu_1 \mu_3 }
g_{\mu_2 \mu_4}  =   g_{\mu_3  [ ax } \, g_{\, bx] \, \mu_4 }} \\
\lefteqn{ \hspace{-0.5cm} 
{\cal F}_{ab}{^{\mu_1}}(x) \, h_{\mu_1 \mu_2 \mu_3 } =
- g_{\mu_2  [\,ax} g_{\, bx] \, \mu_3} + (g_{ax \, bx_{2}} -
g_{ax \, bx_{3}}) g_{\mu_2 \mu_3}}  \nonumber\\
&& \hspace{2.45cm} 
+ {\textstyle{1 \over 2}} g_{ax  \,  bz}  \epsilon^{def}  [g_{\mu_3 \,  dz}
\partial_{a_{2}}  \, g_{ex_{2} \, fz} - g_{\mu_2 \,  dz} 
\partial_{a_{3}}  \, g_{ex_{3} \, fz} ] \label{uno}\\
\lefteqn{ \hspace{-0.5cm} 
{\cal F}_{ab}{^{\mu_1 \mu_2}}(x) \, h_{ \mu_1 \mu_2 \mu_3}
 =   2\, h_{ax \,  bx \, \mu_3}}
\end{eqnarray}

In the last term of equation (\ref{uno}) an  implicit integral
in  $z$  is  assumed.  The  derivatives  that  appear   in    the
above expressions can  be applied over  the  ${\bf R}$'s
integrating by parts, and using the differential  constraint  we
generate from them terms of lower rank in $R$. For example from
(\ref{R1sobreg}) we have
\begin{equation} g_{\mu_2\mu_4}
\partial_{a_{3}} g_{ax \, bx_{3}} R^{(ax, \,bx) \mu_2 \mu_3
\mu_4 } = g_{\mu_2\mu_4} (g_{ax \, bx_{2}} -g_{ax \, bx_{4}}) R^{(ax,
bx) \mu_2  \mu_4 }
\end{equation}
Performing these
calculations, the following partial  results  are obtained for each of
the four expressions quoted above:
\begin{eqnarray*} \lefteqn{
\hspace{-0.7cm} 1)\,  - \epsilon_{abc}  g_{\mu_1 \mu_2} R^{(ax,
bx) \mu_1 \,cx\, \mu_2 } - (g_{ax \, bx_{1}} -g_{ax \,
bx_{2}}) g_{\mu_1\mu_2} R^{(ax, \,bx) \mu_1  \mu_2 }}  \\
\lefteqn{ \hspace{-0.7cm} 2)\;\;\, g_{\mu_1  [ ax } \, g_{\, bx] \,
\mu_2 } R^{(ax, \,bx) \mu_1 \mu_2}} \\ \lefteqn{
\hspace{-0.7cm} 3)\, -  g_{\mu_1  [\,ax} g_{\, bx] \, \mu_2} R^{(ax,
bx) \mu_1 \mu_2 } +   (g_{ax \, bx_{1}} -g_{ax \, bx_{2}})
g_{\mu_1 \mu_2} R^{(ax, \,bx) \mu_1 \mu_2 }} \\ &&
\hspace{4cm} - \epsilon^{def} g_{ax  \,  bz}   g_{\mu_1 \,  dz}  g_{ex
\, fz} R^{(ax, \,bx) \mu_1 } \\ \lefteqn{ \hspace{-0.7cm}
4)\;\;\,    2\, h_{ax \,  bx \, \mu_1} R^{(ax, \,bx) \mu_1 } }
\end{eqnarray*}
Some contributions cancel each other. Integrating by parts the last term
and using the differential constraint we finally obtain
\begin{eqnarray} {\cal  H}(x) \,a_2
({\bf R} )  &=& - 2\; \epsilon_{abc}  g_{\mu_1 \mu_2} R^{(ax,
bx) \mu_1 \,cx\, \mu_2 } \nonumber \\ && + 2\; [ 2\, h_{ax \,
bx \, \mu_1} - \epsilon^{def} g_{ax  \,  bz}   g_{\mu_1 \,  dz}  g_{ex
\, fz}] R^{(ax, \,bx) \mu_1 }  \label{hamrho}
\end{eqnarray}
In the above expression, it can be easily checked that 
the square bracket vanishes identically. Expanding
$R^{(ax, \,bx) \mu_1 \,cx\, \mu_2}$ we get
\begin{equation}
R^{(ax, \,bx) \mu_1 \,cx\, \mu_2} =  -2 \, R^{(ax\,bx\,\mu_1
\,cx\, \mu_{2})_c} + R^{(cx\,ax\,\mu_1 \,bx\, \mu_{2})_c} +
R^{(bx\,cx\,\mu_1 \,ax\, \mu_{2})_c} \label{Rrango5}
\end{equation}
and we therefore see that the contribution of the first term
vanishes due to symmetry
considerations. We conclude that
\begin{equation}
{\cal  H}(x) \,a_2 ({\bf R} )= 0
\end{equation}
We see that the explicit computation of this formal result in the
extended representation involves only a few simple steps. It shows
what a powerful computational tool the use of extended loops provides
us with, just by comparing with the difficult nature of the same
computation done in terms of loops \cite{BrGaPuprl}. The extended loop
analogue of the second coefficient of the Alexander-Conway polynomial
is therefore annihilated formally by all the constraints of quantum
gravity in the extended loop representation.  We will now discuss the
same derivation in a fully regularized context.

\section{Regularized solutions based on extended loops}

As we discussed above, one is forced to regularized the Hamiltonian
constraint to avoid divergences. In order to do that one has to
point-split the expression where repeated spatial points appear in a
multitensor. The regularized expression for the constraint therefore is,
\begin{equation}
{\cal H}^{\,\epsilon} (x) \, \psi({\bf R}) = 
2\int \!d^3 w \!\!\int d^3 u \!\!\int \!d^3 v \,f_{\epsilon} (w,x)
\,f_{\epsilon} (u,x) \,f_{\epsilon} (v,x)\,
\psi ({\cal F}_{ab} (w) \times {\bf   R}^{(au, \,bv)}).
\end{equation}

If one now applies this constraint to the state we considered in the
previous section, one can check that all the computations followed up
(but excluding ) expression (\ref{hamrho}) are well defined and all
cancellations we discussed take place rigorously. The main difference
arises when we integrated by parts the last contribution to
(\ref{R1sobreg}) and applied the differential constraint. Because now
one is dealing with a regularized object, ${\bf R}^{(ax,by)}$ does not
satisfy the differential constraint basepointed at $x$ any longer. The
resulting expression for the action of the constraint is,

\begin{eqnarray}
&& \hspace{-1.9cm}{\cal  H}^{\,\epsilon} (x) \,a_2 ({\bf R} )  =
\int \!d^3 w \!\!\int d^3 u \!\!\int \!d^3 v \,f_{\epsilon} (w,x)
\,f_{\epsilon} (u,x) \,f_{\epsilon} (v,x)\nonumber \\
&&\hspace{-0.8cm}\{- \; \epsilon_{abc}  g_{\mu_1 \mu_2}
R^{(au, \,bv) \mu_1 \,cw\, \mu_2 } \,+\nonumber \\
&& \hspace{0.3cm} [ 2\, h_{aw \,  bw \, \mu_1}
- \epsilon^{def} g_{aw  \,  bz}   g_{\mu_1 \,  dz}  g_{eu \, fz}]
R^{(au, \,bv) \mu_1 } \,+ \nonumber \\
&&\hspace{4cm}
(g_{aw\,bu}-g_{aw\,bv})\,g_{\mu_1 \mu_2} R^{(au\,\mu_1\,bv\,\mu_{2})_c}\}.
\label{hamregrho}
\end{eqnarray}
The first two terms cancel for the same reason a before, but one is left
with the last term (which we call ``anomalous term''). If one studies in
detail the nature of this term when one removes the regulator one finds
a contribution of order $1/\epsilon$,
\begin{equation}
\frac{2}{\sqrt{2\pi}\,\epsilon} \,\epsilon_{abc} \,
g_{\mu_1 \mu_2}\;\partial^{cy}\,
R^{(ax\,\mu_1\,by\,\mu_{2})_c} \mbox{\large{$\mid$}}_{y=x}
\end{equation}
if one assumes that the regulator is Gaussian $f_{\epsilon}
({z})=(\sqrt{\pi} \epsilon )^{-3}\, exp\,(- z^2 \epsilon^{-2})$.

If one renormalizes the Hamiltonian by $\epsilon$ we see that the state
is not annihilated by the constraint. Similar results can be derived for
all other states introduced in section III, the Kauffman bracket and the
exponential of the self-linking number.

Is this the end of the story? Are the states introduced just artifacts
of the formal calculations that do not survive the introduction of a
regularization? We would like to claim that the answer is no. When one
works with regularized expressions one should open up the possibility
that corrections may arise such that in the limit in which regulators
are removed the corrections are zero assuming certain regularity
properties of the wavefunctions. Consider for instance the action of the
operator, 
\begin{equation}
{C}(u,v) = R^{(au \,\uti{\mu} \,bv \,\uti{\nu})_c}
(\frac{\delta}{\delta R^{(aw \, \uti{\mu}  bu) \,\uti{\nu}}} -
\frac{\delta}{\delta R^{(aw \, \uti{\mu}  \,bv)\, \uti{\nu}}})
\end{equation}
on a function of extended loops $\Psi(\bf R)= 
{\bf D}\cdot {\bf R}$ with ${\bf D}$ 
smooth (an example of these kinds of functions would be the holonomy of
a smooth connection). Evidently the action
vanishes due to symmetry arguments in the limit in which $u\rightarrow
v$. However if one considers distributional ${\bf D}'s$ (as we have done
in the explicit 
expressions for the knot 
invariants we propose as possible solutions ) 
the action is not obviously vanishing anymore since in
the limit $u\rightarrow v$ one gets a subtraction of two infinities.
Expressions like this one generate ``anomalous''-like terms,
\begin{equation}
{ C}^{\,\epsilon} (g_{\mu_1 \mu_2} \, R^{\mu_1 \mu_2})=
2 (g_{aw\,bu}-g_{aw\,bv})\,R^{(au \,bv)_c}
\end{equation}

Could it be that adding expressions like the above one to the
Hamiltonian one can cancel the anomalous terms? The answer is in the
affirmative. The precise counter-term is given by the difference of 
two terms, ${\cal C}_2 -{\cal C}_1$, 

\begin{eqnarray}
&&\hspace{-1.72cm}{\cal C}_1 = R^{(au \,\uti{\mu} \,bv \,\uti{\nu})_c}
(\frac{\delta}{\delta R^{aw \, \uti{\mu} \, bu \,\uti{\nu}}} -
\frac{\delta}{\delta R^{aw \, \uti{\mu}  \,bv\, \uti{\nu}}})\\
&&\hspace{-1.72cm}{\cal C}_2 =(
\,R^{(au\, bv) \, \uti{\alpha}}+{\textstyle{\frac{1}{2}}}R^{[au\, bv] \, 
\uti{\alpha}})
(\frac{\delta}{\delta R^{(aw \,  bu)_c \,\uti{\alpha}}} -
\frac{\delta}{\delta R^{(aw \, bv)_c\, \uti{\alpha}}}).
\end{eqnarray}
where the expression $R^{[au\, bv] \, 
\uti{\alpha}}$ is the given by expression (\ref{rsymmetric}) without the 
$(-1)^{n({\uti{\mu}})}$ factor and without the ``rerouting'' action
(the index $\ut{\mu}^{-1}$ is replaced by $\ut{\mu}$. Remarkably,
these expressions also have a simple form in the connection
representation,

\begin{eqnarray*}
&&\hspace{-2cm}{\cal C}_1 :=(A^i_{aw} A^k_{bu} - A^i_{aw} A^k_{bv})
\frac{\delta}{\delta  A^{i}_{au}} \frac{\delta}{\delta A^{k}_{bv}} \\
&&\hspace{-2cm}{\cal C}_2 := (A^i_{aw} A^i_{bu} - A^i_{aw} A^i_{bv})
\frac{\delta}{\delta  A^{k}_{au}} \frac{\delta}{\delta A^{k}_{bv}}.
\end{eqnarray*}

With this {\em single} counter-term all the anomalous contributions to
the action of the Hamiltonian constraint on the $a_2$, the Kauffman
bracket and the exponential of the self-linking number cancel. The
fact that a single counter-term is responsible for all the
cancellations is remarkable and shows that the construction is not
just a gimmick to fix the anomaly problem, but might well be a genuine
counter-term arising from quantum gravity. The fact that the
counter-term has a simple and precise expression in the connection
representation raises the hope that a better intuitive explanation of
it could be gained by viewing it in this context. At present this
issue is not settled: could it be that ${\cal C}_2 -{\cal C}_1$ is
what one needs to add to the Hamiltonian in the connection
representation in order to annihilate the exponential of the
Chern-Simons form when distributional connections are allowed? Could
it reflect the fact that in that case a nontrivial contribution to the
measure arises?  These issues are currently being studied.

\section{Covariance of the extended loop approach}

Due to the results of the previous section, it appears that the extended
representation is a fruitful avenue to regularize quantum gravity. The
fact that quite nontrivial states are found to be regularized solutions to
the constraints is a strong motivation to continue the study of extended
loops. The only obstacle to the use of extended loops as an arena to
quantize gravity seems to stem from the objection that we mentioned at
the beginning concerning convergence issues for the extended holonomy. 

The question of what is a suitable domain in the space of extended loops
to consider for the quantization of gravity so that the covariance of
the formulation under gauge transformations is preserved is a hard
question involving functional spaces. It is by no means settled at
present. What we would like to do in this section is to study several
aspects of these covariance issues. We will conclude it with a
definition of a set of objects based on extended loops for which no
covariance problem arises, to which we can apply the formalism of the
previous sections and which include ordinary loops as a limiting case.

\subsection{The power series holonomy and convergence issues}

In the previous sections we reviewed the construction of a Lie group
structure that includes loops as a particular case, the extended group
of loops.  The main motivation to consider the study of loops has been
the construction of holonomies along the loops given a connection.
Holonomies can be viewed as a Lie group-valued representation of the
group of loops, the Lie group being given by the gauge group.  For the
extended group of loops one can consider similar representations,
typically from the extended group to the enveloping group of the
gauge group (the most general group associated with the dimension of
the representation chosen for the Lie group; for instance, for SU(2)
it would be GL(2,C)). One example of such map, which has been
considered up to present in the literature as the ``extended holonomy''
is simply given by contracting an element of the group of loops,

\begin{equation}
U({\bf X}) = {\bf X} \cdot {\bf A},
\label{xa}
\end{equation}
which formally satisfies $U({\bf X}_1 \times {\bf X}_2) = U({\bf X}_1)
U({\bf X}_2)$.

To distinguish this expression from other possible candidates for a
holonomy we will in this paper call it "power series holonomy".

The above expression implies an infinite summation and its convergence
has to be studied in detail. It should be remarked that this is the case
already for the usual holonomy. The usual holonomy of a connection along
a loop is given by,

\begin{equation}
U_A(\gamma)=\lim_{n\rightarrow\infty} 
\prod_{i=1}^n (1+\delta x^a_i {\bf A}_a)
\end{equation}
and the limit appearing in the product is only well-defined for
particular connections, which we will call "smooth". More precisely, we
will limit ourselves to connections such that the above expression can
be re-expressed as in equation (\ref{xa}) with $X=X(\gamma)$, the
multi-tangents of a loop for all piecewise-smooth loops $\gamma$.

As we noticed, dealing with power-series extended holonomies requires
some detailed understanding of the convergence properties of the
infinite summations involved.  For example, a condition for an
extended holonomy to furnish a correct representation of the extended
group of loops, the holonomy of the product of two extended loops
should equal the product of the holonomies of the individual extended
loops.  This in particular implies that,

\begin{eqnarray}
&&U({\bf X}_1\times{\bf X}_2) = 
\lim_{n\rightarrow \infty} \sum_{k=0}^n
A_{\mu_1,\ldots,\mu_k} (X_1\times X_2)^{\mu_1,\ldots,\mu_k}\\
&&=\lim_{n\rightarrow \infty} \sum_{k=0}^n
\sum_{i=0}^k A_{\mu_1,\ldots,\mu_k} 
X_1^{\mu_1,\ldots,\mu_i} 
X_2^{\mu_{i+1},\ldots,\mu_{k}}  \label{ux1x2}\\
&&=\lim_{{n\rightarrow\infty}\atop {m\rightarrow\infty}} 
\sum_{k=1}^n \sum_{i=1}^m
A_{\mu_{1},\ldots,\mu_{k}}
X_1^{\mu_{1},\ldots,\mu_{k}} 
A_{\nu_{1},\ldots,\nu_{i}}
X_2^{\nu_{1},\ldots,\nu_{i}} \label{ux1ux2}\nonumber\\
&&=U({\bf X}_1)U({\bf X}_2). 
\end{eqnarray}

The equality between (\ref{ux1x2}) and (\ref{ux1ux2}) is not automatic,
one needs certain assumptions about the convergence of the series
involved. For ordinary loops the identity holds, since as we mentioned
we are considering connections such that the usual holonomies furnish a
representation of the group of loops. 

The notion of power-series holonomy we have just defined has the
undesirable property of not being gauge invariant for all elements of
the extended group of loops that satisfy the differential constraint. 
This was discussed in detail in reference \cite{Sch95}.  Basically it was
observed that if one considered an element of the extended group
obtained as the non-rational power of a loop the power-series holonomy
was not gauge invariant.  These explicit examples showed clearly how
dangerous it is to take for granted the convergence of many of the
expressions one considers.  Although formally it is true that if an
extended loop satisfied the differential constraint, the resulting
power-series holonomy as defined above is gauge invariant, in many
practical examples the formal proof fails due to convergence
difficulties. 

There are two different aspects we would like to highlight concerning
these convergence issues. One of them is which are the requirements for
a holonomy of the type ${\bf X} \cdot {\bf A}$ to be well defined and 
under which circumstances {\bf other} definitions of holonomy may be
needed in order to ensure gauge invariance.

\subsection{Examples of extended loops for which the power-series
holonomy is well defined}

When the examples of non-covariant holonomies were first discovered
\cite{Sch95} a concern was raised that maybe the only elements of the
extended group of loops for which the holonomy was well defined were the
ordinary loops. In this subsection we will prove two main results:

a) that there exist many elements of the extended group that are more
general than ordinary loops for which the power-series holonomy is well
defined,

b) that some of these elements form a Lie-group structure.  These
objects will {\bf not} include ordinary loops as a particular case, if
one insists that their power-series holonomy be gauge invariant.  We
will discuss in the following subsection how to define other notions 
of holonomy that would allow us to recover ordinary loops as a
particular case. We will see that these new notions of holonomy can be
viewed as "analytic continuations" of the power-series holonomy outside
of its domain of convergence.

The most trivial example of an element of the group of loops for which
the power-series holonomy is well-defined is furnish by the algebra of
elements first considered by Giles \cite{Gi} which is simply obtained by
linear superpositions of the multi-tangents of ordinary loops.  Given a
finite family $\Gamma$ of $N$ loops $\gamma_i$, one defines a linear
combination

\begin{equation}
X(\Gamma,\vec{\alpha}) =\sum_i^N \alpha_i {\bf X}(\gamma_i),
\end{equation}
and these summations immediately satisfy the differential constraint
since each of the individual ${\bf X}(\gamma_i)$'s does and the
differential constraint is a linear equation. This leads to the
power-series holonomy,

\begin{equation}
U(\Gamma,\vec{\alpha}) = \sum_i^N \alpha_i U(\gamma_i).
\end{equation}

Each of the individual $U$'s in the above summation is an element of
the gauge group, since they are simply holonomies along ordinary
loops.  However, the resulting sum will not be an element of the gauge
group, but of the algebra of $n\times n$ complex matrices if the
representation of the gauge group is  in terms of $n\times n$ matrices.  
This
is not surprising, since one knows that for the power-series holonomy
to belong to the gauge group, the element of the extended group should
satisfy an additional restriction \cite{DiGaGr}, called ``the
algebraic constraint'', and this is a nonlinear equation and therefore
not satisfied in general by the above linear superpositions. Although
one could arrange for some of the examples we will discuss to satisfy
the algebraic constraint, we will for simplicity ignore this
issue. The reason for this is that for the purpose of constructing
extended representations one is mainly interested in producing
holonomies whose trace is gauge invariant and for that it suffices to
satisfy the differential constraint only.

The above set of elements does not in general form a group. The reason
for this is that the elements generated are not in general invertible.
One can gain invertibility by imposing,

\begin{equation}
\sum_{i=1}^N \alpha_i = 1.
\label{condition}
\end{equation}

The inverse of an element of the extended group of loops is given by,

\begin{equation} {\bf X}^{-1} \; = \; X^{-1} {\bf I} \; + \;
\sum^{\infty}_{i=1} (-1)^i X^{-i-1} ({\bf X} \, - \, X {\bf I})^i
\label{inverse} \end{equation} 
and the condition (\ref{condition}),
which is equivalent to $X=1$ assures that the above expression is well
defined.  Notice that the above expression involves only a {\em finite}
sum for each component of the element ${\bf  X}$.  We also see that the
power-series holonomy of such an element involves an infinite expansion
since one adds up all components of the extended element. If one
explicitly computes it, one gets,

\begin{equation} U({\bf X}^{-1}(\Gamma,\vec{\alpha})) \; 
= \; I \; + \;
\sum^{\infty}_{i=1} (-1)^i  (U(\Gamma,\vec{\alpha}) \, - \,  I)^i
\end{equation} 
which is not necessarily convergent for all elements
$X(\Gamma,\vec{\alpha})$. To compute this quantity we have swapped the
order of two infinite summations and we performed first the summation
along all the possible order of components of the multi-tensors in
order to get the holonomy. We will repeat this procedure for other
equations later on.

We will now define a new family of elements which has a well defined
Lie group structure and for which the power-series holonomy is well
defined. Consider the element of the algebra associated to the extended
group of loops,

\begin{equation}
{\bf F}(\gamma,M) \equiv \sum_{j=1}^M {(-1)
 \over j} (I - {\bf X}(\gamma))^j
\end{equation}
and the following associated one parameter family  of elements of the
extended group of loops defined by,

\begin{equation}
{\bf X}(\gamma,M,\lambda) \equiv \exp(\lambda {\bf F}(\gamma,M)).
\end{equation}

The $F$'s and $X$'s satisfy the differential constraint.  The $X$'s are
invertible simply by changing the sign of $\lambda$ and are elements of
the group ${\cal D}_o$ of extended loops that satisfy the differential
constraint base-pointed at $o$ and the $F$'s are elements of its
associated algebra ${d}_o$. 

The motivation for this construction is to notice that in the limit,

\begin{equation}
\lim_{M\rightarrow \infty} {\bf X}(\gamma,M,\lambda) = {\bf X}(\gamma)^\lambda
\end{equation}
the constructed element corresponds to the non-rational power
$\lambda$ of the element $X(\gamma)$ of the group of loops.  We will
not, however, consider this limit in this subsection, since it yields
a non-covariant holonomy.

The elements so introduced obviously form a group for fixed $\gamma$
and $M$ since,

\begin{equation}
{\bf X}(\gamma,M,\lambda_1) {\bf X}(\gamma,M,\lambda_2) =
{\bf X}(\gamma,M,\lambda_1 + \lambda_2) 
\end{equation}

Consider now the contraction of an element of the algebra $F$ with 
a  connection. Such a contraction is well defined in terms of the
holonomies along the loop $\gamma$,

\begin{equation}
{\bf F}(\gamma,M)\cdot {\bf A} = \sum_{j=1}^M {(-1)
\over j} ({\bf I} - {\bf
X}\cdot {\bf A})^j = \sum_{j=1}^M {(-1)
\over j} ({\bf I} - {\bf U})^j.
\label{logarit}
\end{equation}
and is gauge covariant since it is expressed as a finite sum of 
usual holonomies along ordinary loops.

The exponential of the contraction is also well defined in spite of the
fact that it involves an infinite summation since the radius of
convergence of the series expression of the exponential is infinite.
The above expression is gauge covariant and is equal to the 
contraction of a  connection with the exponential of the element
of the algebra,

\begin{equation}
\exp(\lambda {\bf F}(\gamma,M)\cdot {\bf A} ) = (\exp(\lambda 
{\bf F}(\gamma,M)))\cdot {\bf A} = {\bf X}(\gamma,M,\lambda)\cdot{\bf A} 
\end{equation}  

We have therefore explicitly constructed a family of elements of the
group of loops $X(\gamma,M,\lambda)$ such that their power-series
holonomy is well defined. Since all expressions only involve
combinations usual holonomies along ordinary loops they are gauge
covariant under small and large gauge transformations. 

The above family is a one-dimensional Lie group.  One can construct an
infinite dimensional Lie group in a straightforward manner, by
considering the algebra generated by elements ${\bf F}(\gamma,M)$ for
different loops $\gamma$.  Any commutator $[{\bf F}(\gamma_i,M),{\bf
F}(\gamma_j,M)]$ belongs to ${ d}_o$ and has a well-defined
contraction with a connection.

As we mentioned at the beginning, one can recover ordinary loops by
considering the limit $M\rightarrow \infty$. Unfortunately, in that case
it is not true that the power-series holonomy is well defined. If one
considers expression (\ref{logarit}) for that case one has a series 
that does not converge for all values of ${\bf U}$. For instance, for
$SU(2)$ one could consider ${\bf U}= -{\bf I}$ and the series would not
converge. All this is in agreement with the 
results of reference \cite{Sch95}.

\subsection{An analytic extension of the power-series holonomy}

In this subsection we will define a different notion of holonomy, which
can be viewed as an "analytic extension" of the power-series holonomy. 
The construction can be summarized as follows.  We will fist introduce a
family of extended loops satisfying the differential constraint, each
labelled by a loop and a real number, ${\bf X}(\gamma,a)$ such that for
a particular value of $a$ we get the usual multitensor ${\bf
X}(\gamma)$.  We will then consider the power-series holonomy built with
these elements and will show that for certain range of $a$ it is well
defined.  The range will {\em not} include the value for which we
recover the ordinary multi-tensors.  We will then assign a holonomy for
the forbidden range through an analytic extension.  This will provide a
notion of holonomy associated with a family of extended loops in ${\cal
D}_o$ (including ordinary loops) that will coincide with the usual
power-series holonomy only for a certain subset of elements. 

Given a simply-traversed\footnote{Because our main goal is to define
the $\lambda$-th power of a loop with $\lambda$ real it suffices to
consider a simply traversed loop, since a multiply-traversed loop is a
particular case of the construction with $\lambda$ integer.} loop
$\gamma$ and a real number $a$ consider the elements ${\bf
X}(a,\gamma) \in {\cal D}_o$ defined by,

\begin{equation}
{\bf X}(a,\gamma) = (1-a) {\bf I} +a {\bf X}(\gamma).
\end{equation}

We now compute the power-series holonomy for such an element for a
$U(1)$ connection\footnote{The generalization to other gauge groups is
straightforward, one just needs to consider a gauge invariant norm in
the space of matrices of the representation of the given group and
require that the matrices involved be appropriately close to the
identity.}.

\begin{equation}
U({\bf X}(a,\gamma))= {\bf X}(a,\gamma)\cdot {\bf A} = (1-a) +a U(\gamma).
\end{equation}

Since $|U|=1$ then $0<|U({\bf X}(a,\gamma))-1|<2 a$. Then we can
construct an element of the algebra ${d}_o$ defined by

\begin{equation}
{\bf F}(a,\gamma)= \sum_{i=1}^\infty {(-1)\over i} (
1-{\bf X}(a,\gamma))^i
\end{equation}
whose contraction with a connection,
\begin{equation}
{\bf F}(a,\gamma)\cdot{\bf A} = 
\sum_{i=1}^\infty {(-1)\over i} (
1-U({\bf X}(a,\gamma)))^i
\end{equation}
is well-defined and gauge invariant for $a<1/2$ (notice that we have
made a choice in the order in which we perform the two infinite
summations
involved).

We can now introduce a one parameter family of elements of ${\cal D}_o$
obtained by exponentiating the elements of ${ d}_o$ we just
introduced times a real number,

\begin{equation}
{\bf X}(a,\gamma)^\lambda = \exp(\lambda {\bf F}(a,\gamma))
\end{equation}

whose power-series holonomy is well-defined and that can be thought of
as the real powers of the elements ${\bf X}(a,\gamma)$.

Two things need to be noticed.  On the one hand, as we mentioned above,
loops are included as a particular case of this family ($a=1$).  For
that case however, the power-series holonomy for a real power ${\bf
X}(a,\gamma)^\lambda$ is ill defined.  The second observation is that
although the above elements are associated with loops, they do not
provide a representation of the product of loops in the sense that,

\begin{equation}
{\bf X}(a, \gamma_1) \times {\bf X}(a, \gamma_2) \ne 
{\bf X}(a, \gamma_1\circ \gamma_2).
\end{equation}

If one wants to recover loops as a particular case, one needs to
evaluate the above expressions for $a=1$ which is outside the radius of
convergence of the power-series expansions we have considered for the
holonomy. One can proceed by summing the series in the convergence
region to obtain a function of $a$ and then analytically extending such
function to $a=1$. For the $U(1)$ case this can be performed explicitly
in a straightforward fashion. One first observes that in the domain
of convergence of the series,

\begin{equation}
{\bf
F}(a,\gamma)\cdot {\bf A} \equiv \log|U| +i {\rm Arg}(U)
\end{equation}
where ${\rm Arg}(U)$ is the principal part of the argument of the
complex number $U$. This last prescription for the logarithm that is
fixed by the power series expansion we were considering. 

Therefore, exponentiating the above result, we get,

\begin{equation}
U({\bf X}(a,\gamma)^\lambda) = |U({\bf X}(a,\gamma))|^\lambda 
\exp(i \lambda {\rm Arg}(U({\bf X}(a,\gamma)))).
\end{equation}

This expression coincides with the power series expansions we introduced, in
the interval $0 <a<1/2$. We now can analytically continue it to $a=1$ and
obtain as result for the real power of an ordinary loop,

\begin{equation}
U({\bf X}(\gamma)^\lambda) = |U({\bf X}(\gamma))|^\lambda 
\exp(i \lambda {\rm Arg}(U({\bf X}(\gamma)))).
\end{equation}

{}From here one can define a set of extended loops for which their
holonomy is well defined and that includes loops as a particular
case.  Simply consider all the elements obtained by exponentiation of
linear combinations of ${\bf F}(a,\gamma_i)$'s for all possible loops
and their commutators.  We have a consistent prescription to assign
them holonomies given above. It may even be possible that the set
considered above in the case $a=1$ spans the whole $\chi_o$ (the set of
extended loops satisfying both the algebraic and differential
constraint). This is true in the Abelian case, in the non-Abelian case
the issue is more subtle. 

The set we have just constructed can be especially useful to define
"thickened out loops", a procedure we outline in the next subsection.

\subsection{Thickened out loops}

One of the main motivations to consider extended loops was that there exists
a formal set of quite nontrivial solutions to the Wheeler-DeWitt
equation in loop space that is ill defined if one considers it in the
loop representation. By considering its generalization to extended
loops and appropriately restricting the set of extended loops
considered to those that arise from smooth ${\bf Y}$'s the
wavefunctions are well defined. 

Unfortunately the set of extended loops considered faced the
difficulty of yielding non-gauge invariant power-series holonomies.
The attitude that some researchers suggested in the face of this
problem is that probably the extended loops were ``too big'' a set and
that a smaller generalization of loop space would be enough to
regularize the wavefunctions. An intuitive notion that is usually
mentioned in this context is that of ``thickened out loops'' or
``framed loops'', that is, loops with some additional structure that
allows to soften the divergences that the wavefunctions suffer when
evaluated on ordinary distributional loops. 

In this subsection we will show how to implement a practical version
of this notion taking advantage of the constructions we introduced in
the previous section. 

Broadly speaking a ``thickened out loop'' is a tubular structure
composed by a two-parameter family of loops $\gamma(s_1,s_2)$. A naive
way to use such a structure would be to simply superpose the
multi-tangents,

\begin{equation}
{\bf X}_{\rm thick}(\gamma) = \int d^2s {\bf X}_{\rm
thick}(\gamma(s_1,s_2)).
\end{equation}

Unfortunately, such a construction is not very useful in practice. If
one inserts such an expression into one of the wavefunctions
considered, the calculation amounts to contracting it with an
appropriate group of (distributional) propagators. Since the integrand
contracted with the propagators is divergent, the integral is not able
to remove this divergence. Basically what is happening is that one
would like to ``separate'' each entry contracted with a propagator on
different loops. That is not accomplished by the above construction.

One way that seems plausible to achieve that goal is to consider a
superposition similar to the one considered above, but instead of
doing it in terms of elements of the extended group, to do it with
elements of the extended algebra. The result can be exponentiated and
an element of the group can be generated. Concretely,

\begin{eqnarray}
{\bf F}_{\rm thick}(\gamma) &=& \int d^2s {\bf F}_{\rm
thick}(\gamma(s_1,s_2))\\
{\bf X}_{\rm thick}(\gamma) &=& \exp({\bf F}_{\rm thick}(\gamma)).
\end{eqnarray}

Up to present it has not been checked that the above proposal suffices
to yield well-defined expressions for all the wavefunctions involved
in the Kauffman bracket/Jones polynomial construction. It works,
however, for the self-linking number. Consider the expression for the 
self-linking number evaluated on ${\bf X}_{\rm thick}(\gamma)$,

\begin{equation}
{\rm Linking}({\bf X}_{\rm thick}) = 
\left(\exp({\bf F}_{\rm thick}(\gamma))\right)^{\mu_1\,\mu_2}
g_{\mu_1\,\mu_2}.
\end{equation}

If we now write out the expansion of the exponential we get, for the
rank two term,
\begin{equation}
{\rm Linking}({\bf X}_{\rm thick}) = 
\left({\bf F}_{\rm thick}^{\mu_1\,\mu_2}(\gamma) +
{\bf F}_{\rm thick}^{\mu_1} {\bf F}_{\rm thick}^{\mu_2}\right)
g_{\mu_1\,\mu_2}.
\end{equation}

Now, the problematic term would be the first one since it could happen
that both indices are evaluated on the same loop. However that term
vanishes because ${\bf F}$ is an element of the algebra and therefore
satisfies the homogeneous algebraic constraint (it is antisymmetric in
$\mu_1, \mu_2$ whereas $g$ is symmetric). The other term has two
separate thickened out loops and therefore it is naturally
regularized.

A task for the future is to continue this analysis for the other
coefficients of the knot polynomials and see if the prescription
allows to define them without singularities. It will then have to be
studied what is the action of the regularized Wheeler-DeWitt equation
on the resulting states. Notice that the general form of the
constraints in the extended representation is the same as the one one
had in the full extended loops. We are just changing the domain of
dependence from that of extended loops stemming from smooth $Y$'s to
the one we discussed above.

\section{Conclusions and outlook}

We have seen how the use of the Ashtekar variables and the loop
representation in quantum gravity has suggested several heuristic
results concerning the space of states of quantum gravity. In
particular, one is able to find solutions to all the constraints of the
theory (formally speaking) that are related to Chern-Simons theory.
These are the only solutions known at present to all the constraints of
quantum gravity in any formalism. To put such results on a rigorous
setting we need to regularize the theory and we have seen that the use
of extended loops seems a powerful avenue to perform this task. In this
context one can for the first time find well defined solutions
that solve the constraints in a regularized setting. 

The main obstacle at present to consider these results as definitive
is the observation that the use of the full extended group as an arena
may not preserve the gauge covariance of the theory. We have
introduced a sub-sector of the extended space in the previous section
that preserves gauge invariance and that appears as a natural
``thickening out'' regularization for loops. In terms of these objects
it is not clear anymore if the states we have proposed will be
appropriately regularized and if the action of the constraints will be
well defined. These issues are the next step we need to address in the
quest for a regularized theory of quantum gravity.

\end{document}